\documentclass[twocolumn,showpacs,preprintnumbers,amsmath]{revtex4}

\usepackage{graphicx}

\usepackage{amssymb}
\usepackage{dcolumn}
\usepackage{bm}
\usepackage{dsfont}
\usepackage{amsmath}
\usepackage{amsthm}
\usepackage{amscd}

\begin{document}

\title{Dilepton production in $pp$ and $np$ collisions at 1.25 GeV}

\author{B. V. Martemyanov$^{1,2}$, M. I. Krivoruchenko$^{1,2,3}$, and Amand Faessler$^{1}$}
\affiliation{$^{1}$Institut f\"ur Theoretische Physik der Universit\"at T\"ubingen$\mathrm{,}$ Auf der Morgenstelle 14\\ 
D-72076 T\"ubingen, Germany}
\affiliation{$^{2}$Institute for Theoretical and Experimental Physics$\mathrm{,}$ B. Cheremushkinskaya 25\\ 
117259 Moscow, Russia}
\affiliation{$^{3}$Department of Nano-$\mathrm{,}$ Bio-$\mathrm{,}$ Information and Cognitive Technologies\\
Moscow Institute of Physics and Technology$\mathrm{,}$ 9 Institutskii per. \\
141700 Dolgoprudny$\mathrm{,}$ Moscow Region$\mathrm{,}$ Russia}

\begin{abstract}
The inclusive reactions $pp \rightarrow e^+ e^- X$  and $np \rightarrow e^+ e^- X$ 
at the laboratory kinetic energy of 1.25 GeV are investigated in a model of dominance 
of nucleon and $\Delta$ resonances. Experimental data for these reactions have recently been 
reported by the HADES Collaboration. In the original model, the dileptons are produced either from the decays 
of nucleon and $\Delta$ resonances $R \rightarrow N e^+ e^-$ or from the Dalitz decays of $\pi^0$- and $\eta$-mesons created in the $R \to N\pi^0$ and $R \to N\eta$ decays. 
We found that the distribution of dilepton invariant masses in the $pp \rightarrow e^+ e^- X$ reaction is well 
reproduced by the contributions of $R \rightarrow N e^+ e^-$ decays and $R \rightarrow N \pi^0$, $\pi^0 \to \gamma e^+e^-$ decays. 
Among the resonances, the predominant contribution comes from the $\Delta(1232)$, which determines both the direct decay channel 
$R \rightarrow N e^+ e^-$ and the pion decay channel. 
In the collisions $np \rightarrow e^+ e^- X$, 
additional significant contributions arise from the $\eta$-meson Dalitz decays, produced in the $np \rightarrow np\eta$ and  
$np \rightarrow d\eta$ reactions, the radiative capture 
$np \rightarrow d e^+ e^-$, and the $np \rightarrow np e^+ e^-$ bremsstrahlung. These mechanisms may partly explain 
the strong excess of dileptons in the cross section for collisions of $np$ versus $pp$, which ranges from 7 to 100 times 
for the dilepton invariant masses of 0.2 to 0.5 GeV.
\end{abstract}
\maketitle

\section{Introduction}
The HADES Collaboration \cite{HADES} has recently published the results of measurements of dilepton production 
differential cross sections in $pp$ and $dp$ collisions at $T_{LAB} = 1.25$ GeV. Data on the $np$ collisions were 
taken by recording spectator protons from breakup of the deuteron. A huge (from 7 to 100) excess of the dilepton pairs 
in the $np$ collisions as compared to the $pp$ collisions has been observed for the dilepton masses $M = 0.2$ to $0.5$ GeV. 
Conventional sources of dileptons including $\eta$-mesons produced in the $np$ collisions
are not able to explain this excess. 

Bremsstrahlung of dileptons is studied by Kaptari and K\"{a}mpfer \cite{Kaempfer}. The authors conclude that bremsstrahlung of dileptons does not improve 
the agreement: In the $pp$ collisions, the dilepton yield turned out to be overestimated in the entire range of the 
dilepton invariant masses, whereas in the $np$ collisions, 
the dilepton yield was overestimated (underestimated) at lower (higher) values of $M$. Another attempt to describe 
the dilepton data at 1.25 GeV is undertaken by Shyam and Mosel \cite{Shyam}. The authors argue that the inclusion 
of electromagnetic form factors in the bremsstrahlung amplitude of Ref. \cite{Shyam1} practically solves the problem: 
The distribution of invariant masses of dileptons in the $np$ reaction
is found to be consistent with the data. 
\cite{remark}

In this paper the resonance model of Ref. \cite{FFKM}, developed earlier to describe the production of dileptons in 
nucleon-nucleon collisions at $ T_ {LAB} = 1 $ to $6 $ GeV, is used for description of the HADES data.
The basic mechanisms of the dilepton production are supplemented by the 
contribution of bremsstrahlung \cite{Shyam}, scaled by a monopole form factor in the spirit of vector meson dominance 
(VMD) model, and by the radiative capture $np \rightarrow d e^+ e^- $ described on the basis of experimental 
data on deuteron photodisintegration, with the inclusion of the monopole form factor.

\section{Baryon resonance model}

\begin{figure}[t]
\begin{center}
\includegraphics[width=0.382\textwidth]{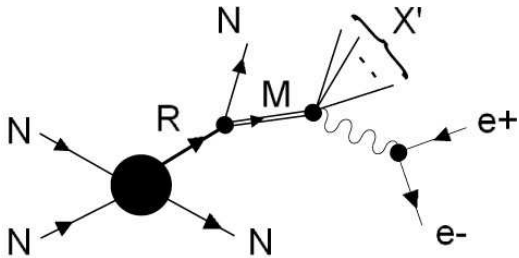}
\end{center}
\caption {
Two-step mechanism for the creation of dilepton pairs in $NN$ collisions. $N$ are nucleons, 
$R$ is nucleon or $\Delta$ resonance, $\mathrm{M}$ is meson arising in the nucleon resonance decay. 
$X^{\prime}$ are mesons and photons. $X^{\prime} = \emptyset$ corresponds to the direct channel 
of vector meson decays $\mathrm{M} \to e^+e^-$.
}
\label{ts}
\end{figure}

In the resonance model, the reaction is a two-stage process. First, nucleon and $\Delta$ resonances are produced in $NN$ collisions, 
at the second step resonances decay into a nucleon, a number of mesons, and a dilepton pair. 
The reaction scheme shown in Fig. \ref{ts}.

For the dilepton production in $p N$ collision at 1.25 GeV we distinguish generally
between three different channels
\begin{eqnarray*}
p N \rightarrow N R &\rightarrow& NN \pi^0,~\pi^0\rightarrow \gamma e^+ e^-,\\
p N \rightarrow N R &\rightarrow& NN \eta,~ \eta\rightarrow \gamma e^+ e^-,\\
p N \rightarrow N R &\rightarrow& NN  e^+ e^-,
\end{eqnarray*}
where $R$ is either a nucleon resonance $N^*$ or a $\Delta$
resonance. The last channel contains all the contributions of the intermediate $\rho$ and $\omega$ mesons.
For the first channel we use here the following isotopic relations for two-nucleon final states:
$pp:pn = 1:1$ for all intermediate resonance $N^* $ except $N^*(1535))$,
$pp:pn = 1:5$ for $N^*(1535))$,
and $p p:p n =1:2$ for $R=\Delta$ \cite{Teis}.
$N^*(1535)$ is the only one resonance giving a significant contribution to the $\eta$-meson production. 
The isotopic ratio for $\eta$, as a result, becomes $p p:p n =1 : 5$ \cite{Teis}. 
The third channel has the same isotopic ratio as the first channel
if one neglects the interference of the intermediate $ \rho$- and $ \omega $-mesons.
This approximation is justified when the contribution of one of the interfering mesons considerably exceeds 
the contribution of the partner. The lack of interference means, in our case, that in the reactions 
$p n \rightarrow p R^0$ and $p n \rightarrow n R^+$
the radiative decays of $ R^0 $ and $ R^+ $ are not coherent and their probabilities are summed up. 
Accordingly, the intermediate vector mesons 
in the $R^0$ and $R^+$ decays do not interfere, although in each of these decays 
the coherence of $ \rho $ and $ \omega $ is preserved.

The Fermi motion of the constituents inside
the deuteron is taken into account using the momentum
distribution of the bound proton extracted from the experimental data 
on the electron scattering cross section on deuteron \cite{fmopid}.

At the  energy of $T_{LAB}=1.25$ GeV the threshold effects for the $\eta$ production become extremely important.
In the $pp$ case, the excess of energy over the threshold in the center-of-mass coordinate system 
equals $\epsilon =-1$ MeV, so the $\eta$-meson is not produced. 
In the $np$ case, the value of $\epsilon$ is positive for both $d\eta$ and $np\eta$ channels.
It is further known 
from the experiment \cite{Calen} that close to the threshold the
$np\rightarrow d\eta$ cross section is greater by a factor of 3 to 4 than
the $np\rightarrow np\eta$ cross section, which in turn is greater than
the $pp\rightarrow pp\eta$ cross section by a factor of 6.5. 
To describe the production of $\eta$-mesons in the resonance models, the channels $pp\rightarrow pp\eta$ 
and $np\rightarrow np\eta$ are usually considered. However, near the threshold it is necessary to take into 
account the channel $np\rightarrow d\eta$, whose cross section is dominant over the others.
We thus add the reaction $np\rightarrow d\eta$ 
to the list of reactions involved in the production of $\eta$-meson.
Below, a parameterization of Ref. \cite{Calen} for the experimental 
$np\rightarrow d\eta$ cross section is used. 
\vspace{5mm}

\section{Bremsstrahlung and radiative capture}

The situation with bremsstrahlung in rather indefinite. Two calculations
\cite{Kaempfer} and \cite{Shyam} that use different vortices for
pion coupling to nucleon give different results. For pseudovector coupling \cite{Kaempfer}
there is an uncertainty of the contact term appearing under the gauging of
pion-nucleon interaction, for pseudoscalar coupling \cite{Shyam} there is an uncertainty of
electromagnetic form factors of photon coupling to pion and nucleons. Nevertheless we will
include bremsstrahlung contribution to our dilepton data. For this we will take the
the results of Ref.\cite{Shyam} available also without electromagnetic form factors and scale 
them by VMD formfactor $1/(1-M^2/m^2_\rho)$.

The radiative capture $np \rightarrow d e^+ e^- $ was never considered as a possible source
of dileptons in $np$ collisions. Unexpectedly its contribution is large in the region 
$M > 0.4$ GeV. To find the cross section of $np \rightarrow d e^+ e^- $ reaction we used
experimental data on deuteron photo-disintegration $\gamma d \rightarrow np$ at photon
energies from 139 to 832 MeV \cite{Dougan}. Kinetic energy of the neutron $T_{LAB}=1.25$ GeV
in the reaction $np \rightarrow d \gamma $ correspond to the photon energy $E_\gamma = 600$ MeV
in deuteron photo-disintegration reaction. The differential cross section of the last
can be parameterized by the linear function \cite{Dougan}
\begin{equation}
\frac{d \sigma}{2\pi d \cos\theta} = 0.6 +a\cdot \cos\theta ~~~~\mu{\rm b/sr}~.
\end{equation}

This gives the total cross section $\sigma(\gamma  d \rightarrow n p) = 7.5 ~~\mu {\rm b}$.
The  cross section of time reversed process $\sigma(np \rightarrow d \gamma)$ is equal to
\begin{eqnarray}
\sigma (np \rightarrow d \gamma) &=& \left(\frac{p^2_{cm}(d\gamma )}{p^2_{cm}(np)}\right)\frac{2\cdot 3}{2\cdot 2}
\cdot \sigma(\gamma  d \rightarrow n p) \nonumber \\
&\approx& 5~~\mu {\rm b}.
\end{eqnarray}

The conversion of the photon to dilepton pair adds to this cross section
phase space correction, form factor (which is taken in the simple VDM form) and
conversion factor
\begin{eqnarray}
\frac{d \sigma(np \rightarrow d e^+ e^-)}{d M} &=& \sigma (np \rightarrow d \gamma)
\frac{p_{cm}(d\gamma^*)}{p_{cm}(d\gamma)} \nonumber \\ 
&\times& \left(\frac{1}{1-M^2/m^2_\rho}\right)^2 \frac{2\alpha}{3\pi M}.
\end{eqnarray}

\begin{widetext}

\begin{figure}[!htb]
\begin{center}

\includegraphics[width=\textwidth]{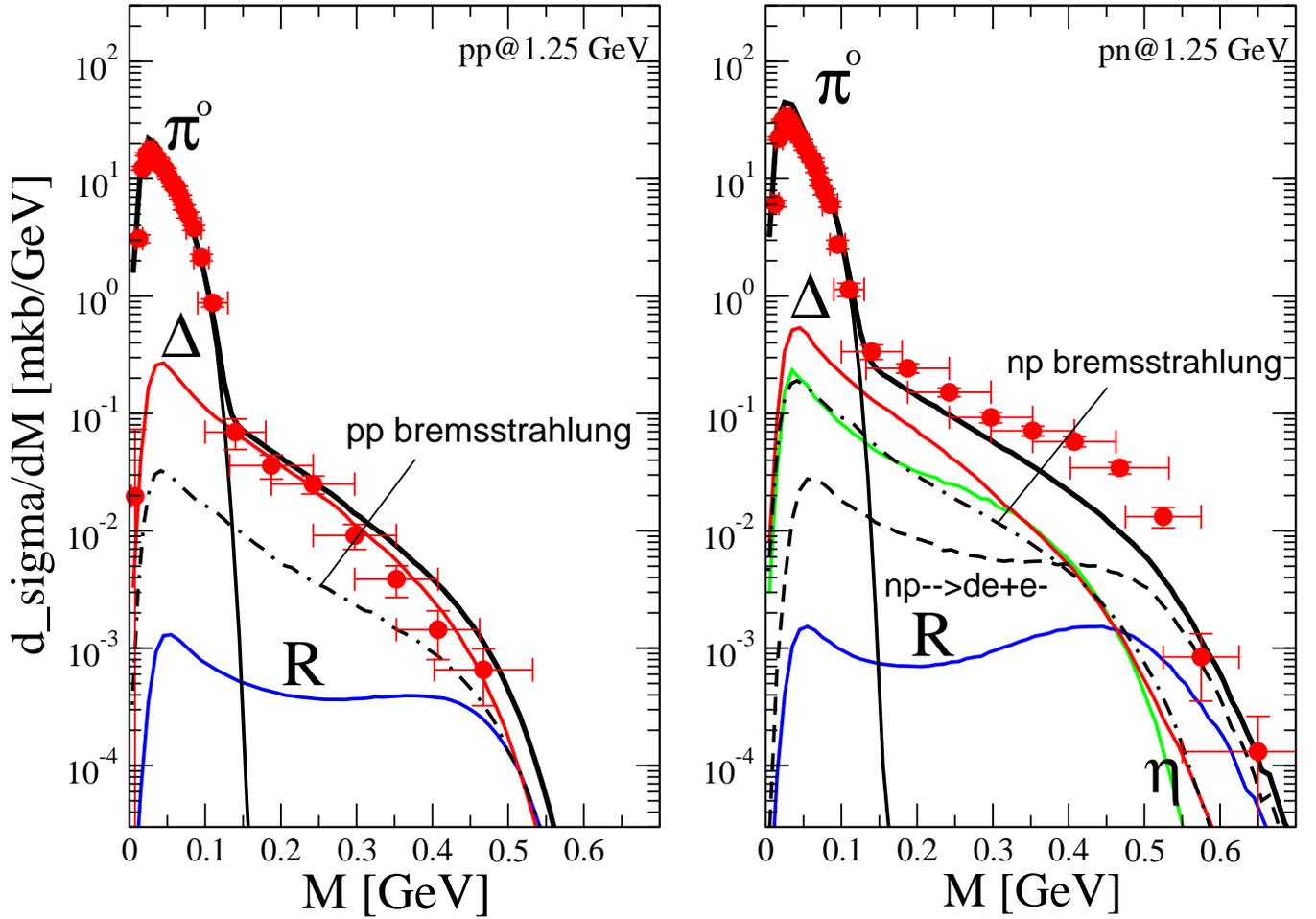}

\end{center}
\caption {(Color online)
Left panel: Dilepton spectrum of the $pp \rightarrow e^+ e^- X$ reaction.
Contributions of $\pi^0$-mesons, $\Delta(1232)$, and other baryon resonances 
are marked by the symbols $\pi^0$, $\Delta$, and $R$, respectively. 
Bremsstrahlung contribution is shown by the dash-dotted line. All contributions are summed up 
incoherently (tick solid line). Right panel: Dilepton spectrum of the $np \rightarrow e^+ e^- X$ reaction.
In comparison to the dilepton spectrum of the $pp \rightarrow e^+ e^- X$ reaction,
additional contributions of the $\eta$-meson Dalitz decay (marked by "$\eta$") and 
the radiative capture $np \rightarrow d e^+ e^- $ (dashed line) are added.
}
\end{figure}

\end{widetext}

\section{Results and discussion}

For comparison with the experimental data, we must take into account the geometrical acceptance of the HADES detector. 
It is used in the form given in the official website of the HADES Collaboration at http://www-hades.gsi.de.

In the case of $pp \rightarrow e^+ e^- X$ reaction (left panel on Fig. 2), 
the dilepton yield is shown separately for the $\Delta(1232)$ Dalitz decay 
(marked by "$\Delta$"), the Dalitz decays of other baryon resonances (marked by "$R$") and
$\pi^0$ (marked by "$\pi^0$"), and for bremsstrahlung (dash-dotted line). 
In the case of $np \rightarrow e^+ e^- X$ reaction (right panel), there are two
additional contributions: the $\eta$-meson Dalitz decay (marked by "$\eta$") and 
the $np \rightarrow d e^+ e^- $ radiative capture (dashed line). 

The results shown in Fig. 2 suggest that the radiative capture is dominated at the dileptons 
invariant masses $M \gtrsim 400$ MeV. There exists the apparent possibility to isolate the contribution 
of other channels by considering the missing mass distribution of the two initial nucleons and the lepton 
pair in the region of $M \gtrsim 400$ MeV. 
The missing mass distribution should reveal a peak corresponding to the deuteron. Quantitative analysis must take 
into account the smearing of the peak associated with the Fermi motion of the proton spectator in the initial-state deuteron.

The excess of the dileptons in comparison with the theoretical estimates at the high invariant masses
indicates a lack of understanding the mechanism of the dilepton pair creation. Among the additional possible sources, 
we wish to point to a dibaryon resonance observed recently in the reaction $ np \to d \pi^0 \pi^0 $
in the energy region of interest \cite{hans}. 

In the model \cite{FFKM}, the dilepton spectrum in the direct channels $\mathrm{M} \to e^+e^-$ is bounded from below 
by two electron masses. In all other models, which we know, there is a cutoff at strong thresholds, i.e., at two 
pion masses for the $\rho \to e^+e^-$ decay, or three pion masses for the $\omega \to e^+e^-$ decay. 
In the dilepton cross section, the strong meson widths are included only in the meson propagators, 
so they do not determine the reaction thresholds. The strong thresholds give rise to certain difficulties in the 
simulation of vector mesons using the Monte Carlo technique. $\rho$-meson, e.g., is generated in accordance with the 
Breit-Wigner distribution
\begin{equation}
dW(\mu )=\frac{1}{\pi }\frac{\mu \Gamma _{\rho}(\mu )d\mu ^{2}}
{(\mu ^{2}-m_{\rho}^{2})^{2}+(\mu \Gamma _{\rho}(\mu ))^{2}},
\label{BW}
\end{equation}
where $\Gamma _{\rho}(\mu )$ is the strong width. When the $\rho$-meson invariant mass $\mu $ approaches 
the two-pion threshold, $\Gamma _{\rho}(\mu )$ decreases, 
the number of the generated $\rho$-mesons decreases, too. However, 
in the dilepton cross section the distribution (\ref{BW}) is then multiplied by the
differential branching ratio of the $\rho \to e^+e^-$ decay, 
\begin{equation}
\frac{d B(\mu ,M)}{dM^2}= \frac{1}{\Gamma _{\rho}(\mu )} \frac{d\Gamma _{\rho \to e^+e^-}(\mu,M )}{dM^2},
\label{BR}
\end{equation}
that contains the strong width in the denominator. 
The result is a numerical instability, although there is a simple cancellation of the two terms.

In order to consistently simulate the subthreshold $\rho$-mesons,
it is sufficient to replace $\Gamma _{\rho}(\mu )$ both in the numerator of Eq. (\ref{BW}) and 
in the denominator of Eq. (\ref{BR}) by any function $f(\mu)$ that is
strictly positive above the two-electron threshold. The cross section does 
not depend on the choice of $f(\mu)$. This feature can be used to check selfconsistency of the code. 

The subthreshold contribution of vector mesons to the dilepton yield 
can be observed experimentally by measuring the exclusive cross section $pp \to ppe^+e^-$. 
Such measurements are planned by the HADES Collaboration. 
Only the direct decay channels $\rho, \omega \to e^+e^-$ contribute to this reaction. 
The experimental data in the subthreshold domain will allow to constrain the 
coupling constants of far off-shell vector mesons with the photon.

\section{Conclusion}
In this paper we showed that the HADES data on the distribution of invariant masses of dileptons 
in the $pp$ collisions at the beam kinetic energy of 1.25 GeV are well reproduced by our resonance model 
and are mainly explained by the $R \to Ne^+e^-$ decays and $\pi^0 \to \gamma e^+e^-$ decays. Moreover, 
the decays of resonances are dominated by $\Delta(1232)$. In the $np$ collisions, an important role 
is attributed to the $\eta$-meson Dalitz decays, 
the radiative capture $ np \to d e^+ e^-$, and the bremsstrahlung. These additional sources provide a partial explanation 
of the observed excess of dileptons in the $np$ collisions.

\acknowledgments
This work is supported by RFBR grant No. 09-02-91341 and DFG grant No. 436 RUS 113/721/0-3. 
B.V.M. and M.I.K. acknowledge the kind hospitality at the University of T\"{u}bingen.


\begin{thebibliography}{99}
\bibitem{remark} {However, complete set of parameters, sufficient for verification of the model predictions, has not been provided.}
\end{thebibliography}
\end{document}